\documentclass[final,3p,times]{elsarticle}

\usepackage{amssymb}
\usepackage{amsmath}

\journal{Journal of Subatomic Particles and Cosmology}

\begin{document}
	
	\begin{frontmatter}
		
		
		
		\title{Fluctuations and Correlations of Conserved Charges in Isobar Collisions at $\sqrt{s_{NN}}$ = 200 GeV with the STAR Detector}
		
		\author[aaa]{Changfeng Li, Hanwen Feng (for the STAR Collaboration)}
		\affiliation[aaa]{organization={Key Laboratory of Quark \& Lepton Physics (MOE) and Institute of Particle Physics, Central China Normal University, 152 Luoyu Road, Wuhan, 430079, Hubei, China}}
		
		\begin{abstract}
			We report measurements of second-order cumulants and correlations of experimentally accessible baryon number, electric charge, and strangeness proxies in Ru+Ru and Zr+Zr collisions at $\sqrt{s_{NN}} = 200$~GeV with the STAR experiment. The results exhibit similar centrality dependences in the two isobar systems within uncertainties. The baryon-strangeness correlation $C_{BS}$ is compared with lattice-QCD and UrQMD calculations. Motivated by recent lattice-QCD predictions for QCD matter in an external magnetic field, several ratios of conserved-charge correlations are studied as functions of centrality. In peripheral collisions, the measured ratios exhibit the same ordering as the lattice-QCD prediction. A qualitatively similar hierarchy is also found in UrQMD, although the model underpredicts the observed peripheral-to-central enhancement. These results motivate further quantitative studies of possible magnetic-field contributions and other sources of the observed centrality dependence.
		\end{abstract}
		
		
		
		
	\end{frontmatter}
	
	
	
	\section{Introduction}
	\label{introduction}
	
	Fluctuations and correlations among conserved charges, such as baryon number ($B$), electric charge ($Q$), and strangeness ($S$), probe the degrees of freedom of strongly interacting matter in relativistic heavy-ion collisions. In particular, the baryon-strangeness correlation was proposed to distinguish between a quark-gluon plasma (QGP) and a hadron gas~\cite{Koch:2005gfr}. If the active degrees of freedom in the QGP are uncorrelated individual quarks, strangeness is carried exclusively by $s$ and $\bar{s}$ quarks, each carrying baryon number $B_s = -S_s/3$, leading to a strong correlation between baryon number and strangeness. In a hadron gas, by contrast, strangeness is carried primarily by kaons, which carry no baryon number, resulting in a weaker correlation. To quantify this difference, the correlation coefficient \(C_{BS}\) is defined~\cite{Koch:2005gfr}. For a gas of uncorrelated quarks, \(C_{BS} = 1\).
	
	More recently, the correlation between baryon number and electric charge, $\chi_{11}^{BQ}$, has been proposed as a probe of magnetic fields in non-central heavy-ion collisions~\cite{Ding:2024mqh}. LQCD computations show that $\chi_{11}^{BQ}$ is strongly sensitive to $eB$~\cite{Ding:2024mqh}, and the magnetic field, mainly generated by spectator protons, is expected to increase toward peripheral collisions~\cite{Ding:2024mqh}. This makes centrality-differential measurements of conserved charge correlations a promising probe of magnetic field effects~\cite{Ding:2024mqh}.
	
	In this paper, we report measurements of correlations and second-order cumulants of conserved charges in Ru+Ru and Zr+Zr collisions at $\sqrt{s_{NN}} = 200$~GeV using the STAR detector. The isobar systems have identical mass numbers but different proton numbers ($Z_{\mathrm{Ru}} = 44$, $Z_{\mathrm{Zr}} = 40$), offering a controlled comparison to study initial-state effects. We present results for $\langle BQ\rangle_c$, $\langle QS\rangle_c$, $\langle BS\rangle_c$, $\langle B^2\rangle_c$, $\langle Q^2\rangle_c$, and $\langle S^2\rangle_c$ as functions of centrality, and compare them with UrQMD calculations.
	
	\section{Experiment and analysis}

	\subsection{Experimental setup and data selection}
	
	The measurements were performed using the STAR detector at RHIC. Charged particles were reconstructed with the Time Projection Chamber (TPC) and identified via ionization energy loss, with the Time-of-Flight detector (TOF) providing additional PID at higher momenta. Centrality is determined using charged-particle multiplicity measured in the Event Plane Detector (EPD), which is separated in pseudorapidity from the particles used in the cumulant analysis, thereby significantly suppressing autocorrelation effects. In this analysis, the phase space is restricted to $0.4<p_T<1.6$~GeV/$c$ and $|y|<0.5$, which ensures good particle identification performance and sufficient statistics for the cumulant measurements.
	
	The analysis uses the 2018 isobar dataset (Ru+Ru and Zr+Zr at $\sqrt{s_{NN}}=200$~GeV) after event quality selections.Proton identification uses TPC alone for $0.4<p_T<0.8$~GeV/$c$ and TPC and TOF for $0.8<p_T<1.6$~GeV/$c$; kaons and pions are identified with TPC and TOF throughout.
	
	$\Lambda$ and $\bar{\Lambda}$ hyperons are electrically neutral and cannot be detected directly; they are therefore reconstructed via their charged decay channel $\Lambda(\bar{\Lambda}) \rightarrow p(\bar{p}) + \pi^-(\pi^+)$ (branching ratio 63.9\%). 
	
	\subsection{Observables}
	
	The correlations and fluctuations are constructed using net $p$, $\Lambda$, $K$, and $\pi$ multiplicities within the measured acceptance as proxies for the conserved charges $B$, $Q$, and $S$. These are experimental proxies constructed within the finite kinematic acceptance and do not represent the full thermodynamic conserved charges.
	
	The net particle numbers are defined as:
	\[
	\Delta p = p - \bar{p},\quad
	\Delta\Lambda = \Lambda - \bar{\Lambda},\quad
	\Delta K = K^+ - K^-,\quad
	\Delta\pi = \pi^+ - \pi^-,
	\]
	with conserved charge proxies defined as
	\[
	B = \Delta p + \Delta\Lambda,\quad
	Q = \Delta p + \Delta K + \Delta\pi,\quad
	S = \Delta K - \Delta\Lambda.
	\]
	
	The second-order cumulants and correlations are:
	\begin{align}
		\langle B^2\rangle_c &= \langle\Delta p^2\rangle_c + \langle\Delta\Lambda^2\rangle_c + 2\langle\Delta p\Delta\Lambda\rangle_c, \\
		\langle Q^2\rangle_c &= \langle\Delta p^2\rangle_c + \langle\Delta K^2\rangle_c + \langle\Delta\pi^2\rangle_c + 2\langle\Delta p\Delta K\rangle_c + 2\langle\Delta p\Delta\pi\rangle_c + 2\langle\Delta K\Delta\pi\rangle_c, \\
		\langle S^2\rangle_c &= \langle\Delta K^2\rangle_c + \langle\Delta\Lambda^2\rangle_c - 2\langle\Delta K\Delta\Lambda\rangle_c, \\
		\langle BQ\rangle_c &= \langle\Delta p^2\rangle_c + \langle\Delta p\Delta K\rangle_c + \langle\Delta p\Delta\pi\rangle_c + \langle\Delta p\Delta\Lambda\rangle_c + \langle\Delta\Lambda\Delta K\rangle_c + \langle\Delta\Lambda\Delta\pi\rangle_c, \\
		\langle QS\rangle_c &= \langle\Delta p\Delta K\rangle_c - \langle\Delta p\Delta\Lambda\rangle_c + \langle\Delta K^2\rangle_c - \langle\Delta K\Delta\Lambda\rangle_c + \langle\Delta\pi\Delta K\rangle_c - \langle\Delta\pi\Delta\Lambda\rangle_c, \\
		\langle BS\rangle_c &= \langle\Delta p\Delta K\rangle_c - \langle\Delta p\Delta\Lambda\rangle_c + \langle\Delta\Lambda\Delta K\rangle_c - \langle\Delta\Lambda^2\rangle_c.
	\end{align}
	Here, $\langle XY\rangle_c \equiv \langle XY\rangle - \langle X\rangle\langle Y\rangle$ denotes the connected correlation function.
	
Several corrections are applied to the raw cumulants. Detector efficiency is corrected using the binomial model with species-dependent average efficiencies \cite{Nonaka:2017, Luo:2019, Chatterjee:2021}. The $\Lambda$ purity correction uses the side-band method, with combinatorial backgrounds estimated from side-band regions via the rotation method \cite{Nonaka:2022}. Feed-down from weak decays of heavier strange hadrons is corrected by replacing $\varepsilon$ with $\varepsilon/(1-f_{\text{feed-down}})$. The centrality bin width correction accounts for volume fluctuations within each centrality bin \cite{Luo:2013}. Statistical uncertainties are estimated via bootstrap, while systematic uncertainties are evaluated by varying track selection, PID cuts, $\Lambda$ reconstruction criteria, and invariant mass fit ranges.

	\section{Results and discussion}
	
\begin{figure}[htbp]
		\centering
		\includegraphics[width=0.51\textwidth]{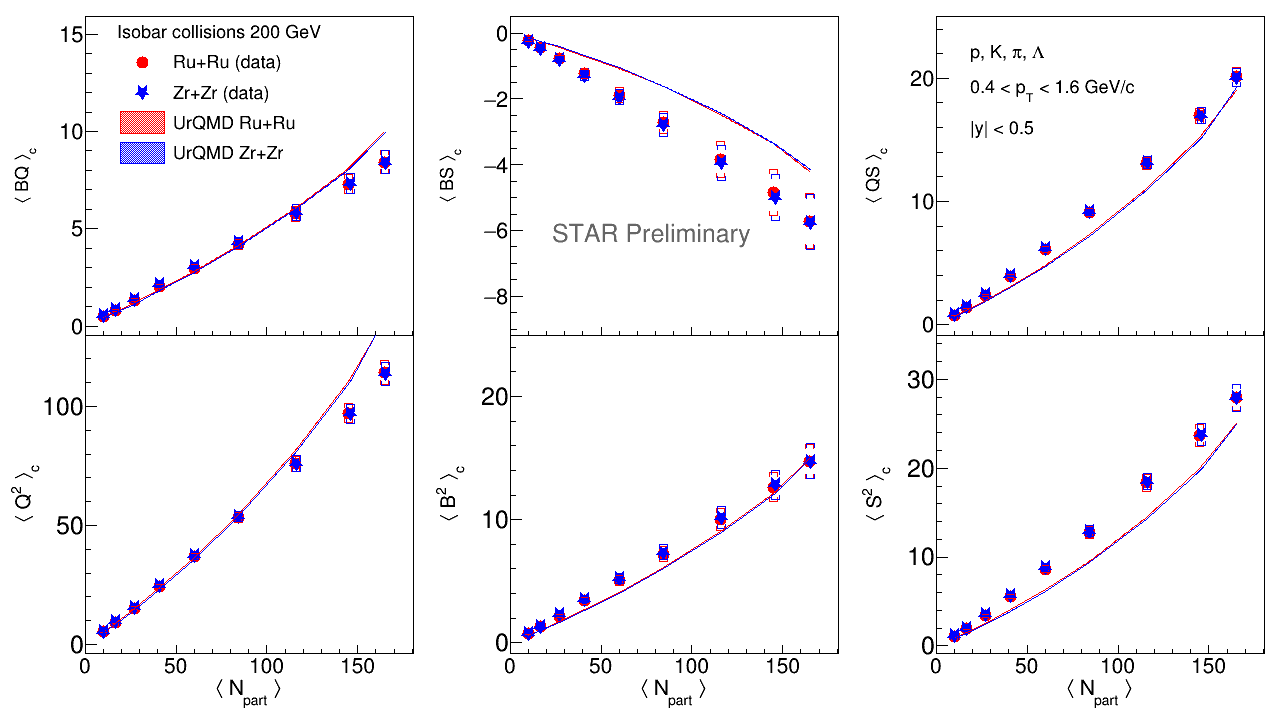}
		\caption{Correlations (upper) and second-order cumulants (lower) as functions of $\langle N_{\mathrm{part}}\rangle$. Red and blue markers represent Ru+Ru and Zr+Zr collisions, respectively.}
		\label{fig:combined_mix}
	\end{figure}

	Figure~\ref{fig:combined_mix} shows the correlations (upper panel) and second-order cumulants (lower panel) as functions of $\langle N_{\mathrm{part}}\rangle$. All quantities increase with $\langle N_{\mathrm{part}}\rangle$, reflecting the larger volume in more central collisions. UrQMD qualitatively describes the trends, and the results are consistent between Ru+Ru and Zr+Zr. Red and blue markers represent Ru+Ru and Zr+Zr collisions, respectively; error bars and brackets represent statistical and systematic uncertainties; bands show UrQMD model calculations.
	
\begin{figure}[htbp]
		\centering
		\includegraphics[width=0.28\textwidth]{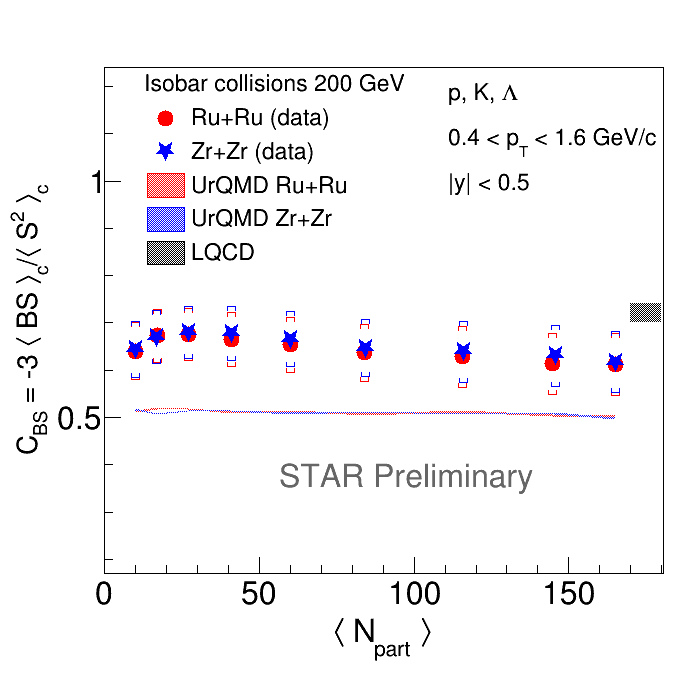}
		\caption{Baryon-strangeness correlation $C_{BS}$ as a function of $\langle N_{\mathrm{part}}\rangle$. The black shaded band is the LQCD result from Ref.~\cite{Bollweg:2024wdf}. The comparison is intended to be qualitative, as the LQCD calculation corresponds to thermodynamic susceptibilities, while the experimental result is obtained from finite-acceptance particle proxies.}
		\label{fig:cbs}
	\end{figure}

	The baryon-strangeness correlation $C_{BS}$ is shown in Fig.~\ref{fig:cbs}. The measured $C_{BS}$ is consistent with both LQCD results \cite{Bollweg:2024wdf} and UrQMD model predictions within uncertainties. It should be noted that while both LQCD and UrQMD are consistent with the data within uncertainties, the two model calculations differ considerably from each other. The difference between the experimental data and the LQCD result may partially arise from the fact that the measured proxies do not fully represent the complete conserved charges of the system, as particles such as $\Xi$ hyperons are not included in the current analysis. No significant dependence on the isobar species is observed within the current uncertainties.
	
\begin{figure}[htbp]
		\centering
		\includegraphics[width=0.7\textwidth]{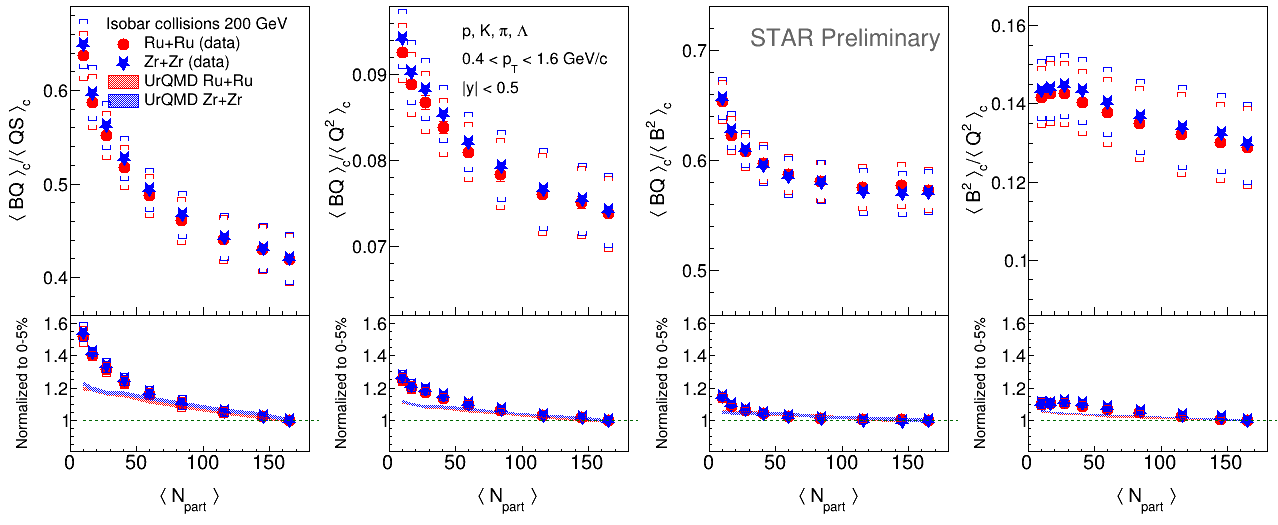}
		\caption{Cumulant ratios (upper) and their ratios normalized to the 0-5\% centrality bin (lower) as functions of $\langle N_{\mathrm{part}}\rangle$.}
		\label{fig:combined_ratio}
	\end{figure}

	Figure~\ref{fig:combined_ratio} shows the cumulant ratios (upper panel) and the same ratios normalized to the 0--5\% centrality bin (lower panel), with the normalized ratios defined as $R_{\mathrm{cp}}(\mathcal{O}) = \mathcal{O}(\text{centrality bin})/\mathcal{O}(0-5\%)$, where $\mathcal{O}$ represents each cumulant ratio. The first three ratios decrease with increasing $\langle N_{\mathrm{part}}\rangle$, showing a strong centrality dependence, while the last ratio exhibits a weaker centrality dependence.
	
	Besides the magnetic field, other quantities such as system volume and particle multiplicity also vary with centrality and may affect the observed ratios. The ratios in Fig.~\ref{fig:combined_ratio} are constructed from cumulants that scale similarly with volume, so volume effects largely cancel. Qualitatively, since the magnetic field is mainly generated by spectator protons, its strength is expected to increase from central to peripheral collisions.
	
	LQCD predicts a hierarchy of conserved charge correlations, $\langle BQ\rangle_c/\langle QS\rangle_c > \langle BQ\rangle_c/\langle Q^2\rangle_c > \langle BQ\rangle_c/\langle B^2\rangle_c > \langle B^2\rangle_c/\langle Q^2\rangle_c$, in the presence of magnetic fields \cite{Ding:2025abc}. We observe the same ordering in peripheral Ru+Ru and Zr+Zr collisions. Interestingly, UrQMD, which does not incorporate electromagnetic-field effects, also reproduces the qualitative ordering of these ratios. Therefore, the hierarchy itself cannot be regarded as an unambiguous signature of magnetic fields. The stronger peripheral-to-central enhancement observed in the data compared with UrQMD may indicate additional centrality-dependent physics not captured by the model. Whether this difference can be attributed to magnetic-field effects requires further quantitative investigation. No significant difference between the Ru+Ru and Zr+Zr systems is observed within the present uncertainties. Given the relatively modest difference in the expected magnetic-field strength between the two isobar systems, the current precision may not be sufficient to resolve such an effect.
	
	\section{Summary}	
	We have measured correlations and second-order cumulants of conserved charges in Ru+Ru and Zr+Zr collisions at $\sqrt{s_{NN}} = 200$~GeV. Results are consistent between the two isobar systems. The baryon-strangeness correlation $C_{BS}$ is consistent with both LQCD and UrQMD within uncertainties. LQCD predicts a hierarchy, $\langle BQ\rangle_c/\langle QS\rangle_c > \langle BQ\rangle_c/\langle Q^2\rangle_c > \langle BQ\rangle_c/\langle B^2\rangle_c > \langle B^2\rangle_c/\langle Q^2\rangle_c$, in magnetic fields, and we observe the same ordering in peripheral collisions. However, UrQMD without magnetic fields reproduces a qualitatively similar ordering, and no significant Ru+Ru--Zr+Zr difference is seen. Thus, the ordering alone does not establish a magnetic-field effect. The stronger peripheral-to-central enhancement in data compared to UrQMD suggests additional centrality-dependent physics beyond the model, possibly including magnetic-field effects, but requires further investigation.
	
	Measurements at additional beam energies, together with higher-order conserved-charge correlations and quantitative calculations incorporating realistic electromagnetic-field evolution, will be essential for disentangling possible magnetic-field effects from conventional sources of centrality dependence.
	
	\section*{Acknowledgments}
	
	This work is supported in part by the National Natural Science Foundation of China under Grant Nos. 12325508 and 12293064, the National Key Research and Development Program of China under Contract No. 2022YFA1604900, and the Fundamental Research Funds for the Central Universities.

	\bibliographystyle{elsarticle-num}
	\bibliography{sqm2026_template}
	
	
	
\end{document}